\documentclass{article}
\usepackage[utf8]{inputenc}
\usepackage{times}
\usepackage{amsfonts}
\usepackage{amsmath,amssymb}
\usepackage{indentfirst}
\usepackage{wrapfig}
\usepackage{graphicx}
\usepackage{graphics}
\usepackage{caption}
\usepackage{subcaption}
\usepackage[toc,page]{appendix}
\usepackage{color}
\usepackage{cite}
\usepackage[inactive]{srcltx}
\usepackage[T1]{fontenc}
\usepackage{float}
\usepackage{hyperref}

\begin{document}
	\date{}
\begin{center}
	{\Large\bf A note on the emission spectrum and trapping states in the Jaynes-Cummings model}
\end{center}
\begin{center}
	{\normalsize J.L.T. Bertassoli and A. Vidiella-Barranco \footnote{vidiella@unicamp.br}}
\end{center}
\begin{center}
	{\normalsize{Gleb Wataghin Institute of Physics - University of Campinas}}\\
	{\normalsize{ 13083-859   Campinas,  SP,  Brazil}}\\
\end{center}
\begin{abstract}
The emission of light from an atom represents a fundamental process that provides valuable insights into the atom-light interaction. The Jaynes-Cummings model is one of the simplest fully quantized models to deal with these interactions, allowing for an analytical solution, while exhibiting notable non-trivial effects. We explore new features in the fluorescence emission spectrum for initial ``trapping states'', which suppress the atomic population inversion. Despite the seemingly dormant activity of the atom, the resulting emission spectra exhibit rich features, and using a dressed-state coordinates formalism, we are able to quantitatively explain the different profiles in the spectrum. We generalize the trapping conditions for non-zero atom-field detuning and also unveil two types of trapping states that lead to spectra with three peaks, in contrast to previously known states: a center peak and one secondary peak on each side. These are a trapping state formed by a Schrödinger cat state with Poissonian statistics (Yurke-Stoler state) and also a different type of ``perfect trapping state''. 
\end{abstract}
%
\section{Introduction}
The Jaynes-Cummings model, presented in 1963 \cite{cummings63}, is the very basic quantum model to describe the light-matter interaction: a single two-level atom coupled, near resonance, to a single mode of the quantized electromagnetic field. It can be solved analytically under the rotating-wave approximation, and despite its simplicity, this model has revealed numerous interesting and non-trivial quantum effects. Moreover, it has been instrumental in understanding and conducting fundamental experiments, especially (but not only) in the field of cavity quantum electrodynamics (QED). We can find comprehensive reviews about this model in the literature, e.g., in \cite{knight93}, and  \cite{larson21} for a more recent account. Here we are interested in uncovering novel effects in the Jaynes-Cummings using the general approach developed in \cite{vidiella99}. We study for that certain features of the light emitted by the two-level atom are affected if we consider specific initial conditions, namely product states of the atom-field system known as trapping states. It was pointed out in \cite{zubairy89} that the atomic population inversion can be strongly suppressed if the atom is initially prepared in a coherent superposition of its lower and upper states and the field in a coherent state. This phenomenon, known as atomic population trapping, may also arise for other initial product states, as will be shown in what follows. A natural basis to address such process is the one formed by the eigenstates of the JCM, the ``dressed-states'', as discussed in \cite{cirac91}. Furthermore, in \cite{vidiella99} it was given a quantitative explanation of the population trapping using a representation of the atom-field state in terms of a set of parameters, the ``dressed-state coordinates'', also used to study the population trapping in a mazer \cite{solano03}. This allowed us to naturally consider the joint initial properties of the atom-field system as well as to find simple analytical expressions for the atomic populations that exhibit the conditions needed for the trapping. 

One way of retrieving relevant information from the atom-field system is by analysing the spectral features of the light emitted by the atom. An intuitive appealing definition of emission spectrum was given in \cite{eberly77}, where the authors propose a time-dependent ``physical spectrum'' of resonance fluorescence having a direct connection with experimental observations. This is a widely used methodology to investigate the emission spectrum in the Jaynes-Cummings model \cite{eberly80}, and it will also be suitable for our purposes. Here, similarly to the approach used to study the atomic population inversion \cite{vidiella99} for trapping states, we will also express the emission spectrum in terms of the dressed-state coordinates, which allows a quantitative assessment of the peaks' forms in the emission spectrum. This is a peculiar scenario, where considering initial atom-field trapping states, a relevant aspect of the atomic dynamics (population inversion) is suppressed at all times, but the atomic dipole activity is evidenced by the (experimentally accessible) spectrum of light emitted by the atom. For instance, while the atomic inversion can be identically zero for different initial states (perfect trapping states), the corresponding emission spectrum can show different patterns, making it better suited to analyse some properties of the JCM. We have identified different types of trapping states, one of them being a perfect trapping state, and the other comprising a Schr\"odinger cat state of the field having Poissonian statistics (Yurke-Stoler state). Also, we have considered the off-resonant case (non-zero atom-field detuning), for which we have found the condition for trapping and also verified subtle differences in the profiles of the emission spectrum.   

This paper is organized as follows: in Section 2, we review the basics of the Jaynes-Cummings model and introduce the dressed-state coordinate formalism; in Section 3, we review the spectrum of resonance fluorescence for general initial conditions. In Section 4 we investigate the spectrum of different types of trapping states using the dressed-state coordinate formalism, including novel types of trapping states and also show the effects of a non-zero atom-field detuning. In Section 5 we present our conclusions.

\section{Dressed-state coordinates in the Jaynes-Cummings model\label{section2}}

We consider a two-level atom with lower (upper) energy state $|g\rangle$ $\left(|e\rangle\right)$ interacting with a single-mode of the quantized electromagnetic field under the rotating-wave approximation (Jaynes-Cummings model). The Hamiltonian for this model, disregarding the vacuum energy of the field, is \cite{cummings63}
\begin{equation}
	\hat{H} = \frac{1}{2}\hbar{\omega}_{a}\hat{\sigma}_{z}+\hbar\omega_f\hat{a}^{\dagger}\hat{a}+\hbar \lambda \left( \hat{\sigma}_{+}\hat{a}+\hat{\sigma}_{-}\hat{ a}^{\dagger}\right)  \label{jcmhamiltonian}.
\end{equation}
Here $\omega _{a}$ is the atomic transition frequency, $\omega _{f\text{ }}$ is the frequency of the cavity field, $\lambda $ is the atom-field dipole coupling constant, $\hat{a}^{\dagger }(\hat{a})$ is the photon creation (annihilation) operator of the field and $\hat{\sigma}_{z} = |e\rangle\langle e| - |g\rangle\langle g|,\,\hat{\sigma}_{+} = |e\rangle\langle g|,\,\hat{\sigma}_{-} = |g\rangle\langle e|$ are the atomic inversion, rising and lowering operators. We denote the atom-field detuning, the difference between the atomic and the cavity field frequencies, as  $\Delta = \omega_a - \omega_f$.

The eigenstates of the interaction picture Hamiltonian $\hat{H}_I = \hbar \lambda \left( \hat{\sigma}_{+}\hat{a}+\hat{\sigma}_{-}\hat{ a}^{\dagger}\right)$, known as dressed states, are the superposition states $|n\pm\rangle$, given by
\begin{eqnarray}
	&&|n+\rangle = \cos{\beta_n}|n,e\rangle + \sin{\beta_n}|n+1,g\rangle \\ \nonumber
	&&|n-\rangle = -\sin{\beta_n}|n,e\rangle + \cos{\beta_n}|n+1,g\rangle,
\end{eqnarray}
where $|n\rangle$ is the Fock state of the field with $n$ photons, $\hat{a}^{\dagger}\hat{a}|n\rangle = n|n\rangle$. The angles $\beta_n$ are
\begin{equation}
	\label{beta_n}
	\beta_n = \arctan\left( \frac{2\lambda\sqrt{n + 1}}{\Omega_n + \Delta}\right),
\end{equation}
and $\Omega_n$ is the Rabi frequency
\begin{equation}
	\Omega_n = \sqrt{\Delta^2 + 4\lambda^2(n+1)}.
\end{equation}
The corresponding eigenenergies are $E_{n\pm} = \pm \frac{\hbar}{2}\Omega_n$.

It can be enlightening to expand an arbitrary initial atom-field state in terms of dressed states. In \cite{vidiella99} a set of suitable ``dressed-state coordinates'', namely $w_{n},\,\chi _{n}$, and $\theta_{n},\,\phi _{n}$ was introduced, with $w_{n}\in \left[ 0,1\right]$ $\left(\sum_{n=-1}^{\infty }w_{n}^{2}=1\right)$, $\theta_{n}\in \left[ 0,\pi \right]$ and $\chi _{n},\phi _{n}\in \left[ 0,2\pi \right]$. In particular, any initial atom-field state\footnote{We restrict our analysis to pure states.} can be expanded in terms of the dressed states $|n\pm\rangle$, and expressed in terms of the dressed-state coordinates as:
\begin{equation}
	|\Psi(0)\rangle = w_{-1}|0,g\rangle+\sum_{n=0}^{\infty} w_{n} \exp\left(i\chi _{n}\right)|\psi_n\rangle,
\end{equation}
with 
\begin{equation}
	|\psi_n\rangle = \left[\cos\left(\frac{\theta_{n}}{2}\right)|n+\rangle + e^{-i\phi _{n}}\sin\left(\frac{\theta_{n}}{2}\right)|n-\rangle \right].
\end{equation}
The coordinate $\theta_n$ gives a measure of the degree of proximity to the nearest dressed state, the ``dressedness'', while the weights $w_n$ measure the relative importance of the component states $|\psi_n\rangle$ in $|\Psi(0)\rangle$. Also, $w_n^2$ is related to the probability distribution of the total excitation number operator $\hat{N} = \frac{1}{2}\hat{\sigma}_z + \hat{a}^{\dagger}\hat{a}$. For instance, We note that, in the special case where the atom is initially prepared in one of the states $\left| e\right\rangle $ or $\left| g\right\rangle $, then $\theta _n\equiv \frac \pi 2$ and $\phi_n\equiv 0\left( \pi \right) $ for all $n$, while $w_n^2$ reduces to the photon number distribution: 
\begin{equation}
	w_n^2\rightarrow \left\{{P_n=\left| \left\langle n|\Phi \left( 0\right) \right\rangle _f\right| ^2\text{ (for }\left| e\right\rangle ) \atop P_{n+1}=\left| \left\langle n+1|\Phi \left( 0\right) \right\rangle _f\right| ^2\text{ (for }\left| g\right\rangle )}\right. ,
\end{equation}
being $|\Phi(0)\rangle_f = \sum_{n=0}^\infty a_n |n\rangle$ the initial state of the field. The evolution of the dressed-state coordinates in the Jaynes-Cummings model is such that $w_{n}$ and $\theta_{n}$ are constants of motion while $\chi _{n}$ and $\phi _{n}$ have a simple time-dependence \cite{vidiella99}. 

The atomic inversion for a general initial condition can be written in a relatively compact form,
\begin{equation}
	\left\langle \hat{\sigma}_z \right \rangle = -\omega_{-1}^2 + \Delta \sum_{n=0}^{\infty} \frac{w_n^2\cos\theta_n}{\Omega_n} + \sum_{n=0}^\infty \frac{2\lambda\sqrt{n+1}}{\Omega_n}w_n^2\sin\theta_n\cos\phi_n(t).\label{inversiondressed}
\end{equation}
A particularly relevant quantity appearing in Eq.(\ref{inversiondressed}) and introduced in \cite{vidiella99} is the ``weighted dressedness distribution''
\begin{equation}
	D_n = w_n^2\sin\theta_n.\label{dressedness}
\end{equation}
As seen in Eq.(\ref{inversiondressed}), for general initial conditions, it is $D_n$, and not the photon number distribution $P_n$ that governs the evolution of the atomic inversion. More precisely, if a component $|\psi_n\rangle$ of a given initial state $|\Psi(0)$ has a large ``dressedness'', that is, $\sin(\theta_n) \rightarrow 0$, it will contribute very little to the overall inversion. Besides, even if $|\psi_n\rangle$ allows a large inversion, this may not be relevant to the resulting atomic inversion if the weight factors $w_n^2$ (relative importance) are small. We therefore conclude that initial states having a large ``dressedness'' lead to a virtually constant value for the atomic inversion, $\left\langle \hat{\sigma}_z \right \rangle \approx -\omega_{-1}^2 + \Delta \sum_{n=0}^{\infty} \left(w_n^2\cos\theta_n\right)/\Omega_n$.
If we assume an initial atom-field state of the form
\begin{equation}
	|\Psi\left(0\right)\rangle = \left(\cos \varphi |e\rangle + \text{e}^{-i\eta} \sin \varphi |g \rangle \right) |\alpha\rangle,\label{zaheerstate}
\end{equation}
where $|\alpha\rangle$ is a coherent state with $\alpha = |\alpha| \text{e}^{i\nu_\alpha}$, we obtain
\begin{equation}
	w_n^2 = \frac{|\alpha|^{2n}\text{e}^{-|\alpha|^2}}{(n+1)!} \left[ (n+1)\cos^2\varphi + |\alpha|^2\sin^2\varphi \right],
\end{equation}
and
\begin{equation}
	D_n = \frac{|\alpha|^{2n}\text{e}^{-|\alpha|^2}}{(n+1)!} \frac{2\lambda\sqrt{n+1}}{\Omega_n} \bigg| (n+1)\cos^2\varphi - |\alpha|^2 \sin^2 \varphi \,\text{e}^{2i(\nu_\alpha-\eta)}- \frac{\Delta}{2\lambda}|\alpha|\sin(2\varphi)\text{e}^{i(\nu_\alpha-\eta)}\bigg|.
\end{equation}
The weight factors $w_n^2$ follow a (coherent state) Poisson distribution when $\varphi = 0$ and have small deviations from that distribution for any value of $\varphi$. Thus, $w_n^2$ is maximized around the integer $n_\text{max}$ closest either to $|\alpha|^2$ or $|\alpha|^2 - 1$. Also, the $\sin\theta_n=D_n/w_n^2$ distribution has an absolute minimum, located around $n_\text{min} \approx |\alpha|^2 \tan^2 \varphi + \frac{\Delta}{\lambda}|\alpha| \tan \varphi - 1$, with minimum value given by $\sin\theta_{n_\text{min}}\approx|\sin(\nu_\alpha - \eta)|$. This means that the atomic population inversion will be quenched if the relative phase in the initial atomic state matches the phase of the initial field state, i.e, $\eta = \nu_\alpha$ (or $\eta = \pi + \nu_\alpha$), and the minimum of the $\sin\theta_n$ distribution matches the maximum of $w_n$, i.e., $n_\text{max} = n_\text{min}$, what gives the condition $\tan(2\varphi)\approx \frac{2\lambda|\alpha|}{\Delta}$.

\section{The resonance fluorescence spectrum}

Resonance fluorescence is a process when light is emitted after an atomic system (a two-level atom in this case) is excited by incident light. The emitted light can hold interesting properties of the incident light as well as information about the fluorescence process itself. Interestingly, a system as simple as a two-level atom can reveal a rich variety of details in its emission spectrum. One way of characterizing the emitted light, is to use a frequency-sensitive filter to analyse the outgoing light. In this setup, the spectrum is essentially the intensity of the emitted light crossing the filter as a function of the frequency setting of the filter. In this spirit, it can be defined \cite{eberly77} a suitable ``physical spectrum'' $S(\delta)$ given by
\begin{equation}
	S(\delta) = \operatorname{Re} \left\{ \int_0^\infty \exp[-i(\lambda\delta + \omega_f)\tau] \exp(-\gamma\tau) \overline{\Gamma}(\tau) d\tau \right\}.
\end{equation}
Here $\gamma$ is the bandwidth of the detector, and $\delta = (\omega - \omega_f)/\lambda$, where $\omega$ is the frequency at which the measurement is taking place. The function $\overline{\Gamma}(\tau)$ is the time average (over a period $T$) of the two-time, dipole-dipole correlation function. In other words,
\begin{equation}
	\overline{\Gamma}(\tau) = \frac{1}{T} \int_0^T \Gamma(t,\tau) dt, 
\end{equation}
where
\begin{equation}
	\Gamma(t,\tau) = \langle\hat{\sigma}_+(t+\tau)\hat{\sigma}_-(t)\rangle.
\end{equation}
We have thus in hands the physical spectrum, basically the Fourier transform of the dipole-dipole correlation function of the atom. Within the all-quantized scenario offered by the Jaynes-Cummings model, one may investigate the possible effects when considering different initial states of the atom-field system. For instance, as discussed in \cite{eberly83}, an initial coherent state of the field $|\Phi(0)\rangle_f = |\alpha\rangle$ already brings interesting phenomena. If $\alpha = 0$ (initial vacuum state $|0\rangle$) and having the atom initially in the excited state $|\varphi(0)\rangle_a = |e\rangle$, the ``vacuum Rabi splitting'' occurs. This peculiar phenomenon, taking place in the absence of cavity photons, consists in the formation of two peaks in the emission spectrum centered at the frequencies $\omega = \omega_f \pm \lambda$, being a signature of the vacuum Rabi oscillations.  However, by increasing the coherent amplitude $\alpha$, the vacuum splitting progressively disappears, and the spectrum changes towards the three-peaked structure analogous to the Mollow triplet \cite{mollow69} which arises if the two-level atom interacts with a sufficiently intense classical field. More generally, we may calculate the emission spectrum for an initial state of the type $|\Psi(0)\rangle = |\varphi(0)\rangle_a\otimes|\Phi(0)\rangle_f$, where
\begin{equation}
	\label{generalstate}
	|\varphi(0)\rangle_a = \left(\cos{\varphi}|e\rangle + e^{-i\eta}\sin{\varphi}|g\rangle\right), \ \ \mbox{and} \ \ |\Phi(0)\rangle_f = \sum_{n=0}^\infty a_n |n\rangle.
\end{equation} 
\begin{eqnarray}
	\label{spectrumbare}
	S(\delta) = \sum_{n=0}^{\infty} \left[ \frac{\gamma \sin^2\beta_{n-1}}{\gamma^2 + \lambda^2(\delta - (\zeta_n - \zeta_{n-1}))^2} + \frac{\gamma \cos^2\beta_{n-1}}{\gamma^2 + \lambda^2(\delta - (\zeta_n + \zeta_{n-1}))^2}  \right] \bigg[ |a_n|^2\cos^2\varphi\cos^4\beta_n \\ \nonumber
	+ |a_{n+1}|^2\sin^2\varphi\sin^2\beta_n\cos^2\beta_n
	+ 2\operatorname{Re} (e^{i\eta} a_n a_{n+1}^*) \sin\varphi\cos\varphi \sin\beta_n\cos^3\beta_n \bigg] \\ \nonumber
	+ \left[ \frac{\gamma\sin^2\beta_{n-1}}{\gamma^2 + \lambda^2(\delta + (\zeta_n + \zeta_{n-1}))^2} +  \frac{\gamma\cos^2\beta_{n-1}}{\gamma^2 + \lambda^2(\delta + (\zeta_n - \zeta_{n-1}))^2} \right] \bigg[ |a_n|^2\cos^2\varphi\sin^4\beta_n \\ \nonumber
	+ |a_{n+1}|^2\sin^2\varphi\sin^2\beta_n\cos^2\beta_n
	- 2\operatorname{Re} (e^{i\eta} a_n a_{n+1}^*) \sin\varphi\cos\varphi \sin^3\beta_n\cos\beta_n \bigg],
\end{eqnarray}
where 
\begin{equation}
	\zeta_n = \sqrt{\left(\frac{\Delta}{2\lambda}\right)^2+n+1}.
\end{equation}
In Eq.(\ref{spectrumbare}) $\beta_n$ was extended so that $\beta_{-1} = 0$ even when Eq.(\ref{beta_n}) is undefined at $\Delta = 0$.

\begin{figure}[ht!]
	\centering\includegraphics[width=9cm]{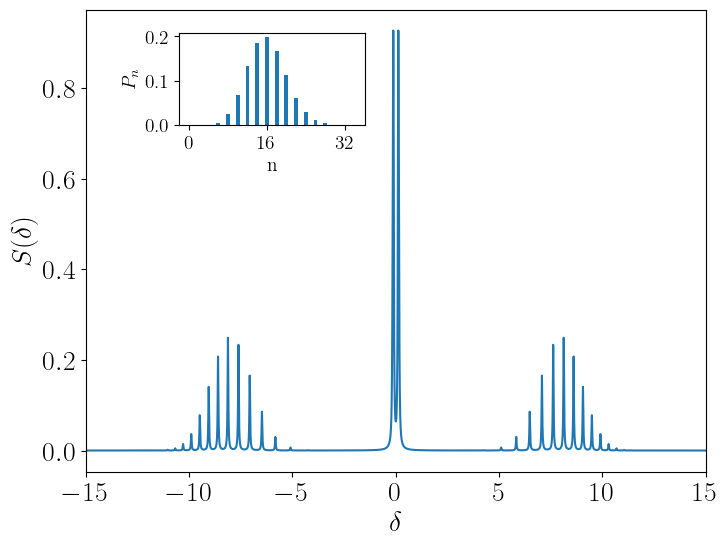}
	\caption{Emission spectrum for the atom initially in its upper state $|e\rangle$ and the field in an even coherent state with $|\alpha| = 4.0$, and $\Delta=0$. The photon number distribution of the initial field state is plotted in the inset.}\label{fig:spectrumcat}
\end{figure}

\begin{figure}[ht!]
	\centering\includegraphics[width=9cm]{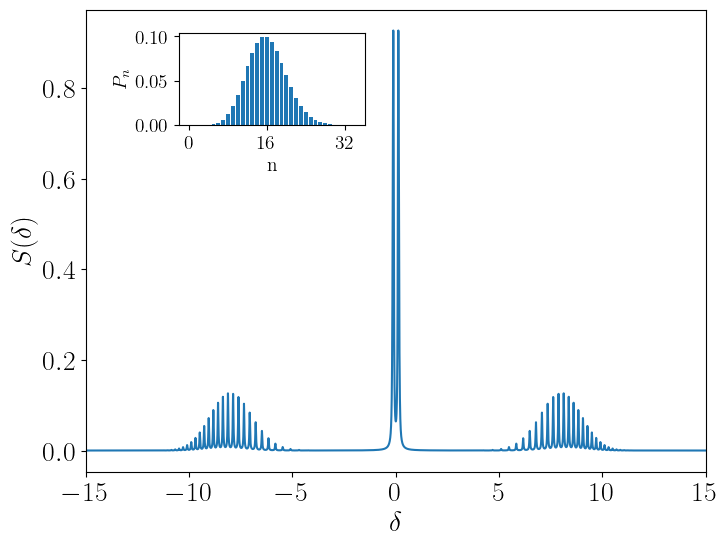}
	\caption{Emission spectrum for the atom initially in its upper state $|e\rangle$ and the field in a coherent state with $|\alpha| = 4.0$, and $\Delta=0$. The photon number distribution of the initial field state (Poisson distribution) is plotted in the inset.}\label{fig:spectrumcoherent}
\end{figure}

In \cite{moyacessa20}, the authors consider the field initially in a squeezed coherent state $|\Phi(0)\rangle_f = |\xi,\alpha\rangle$, a state characterized by a squeezing parameter $\xi$ and a coherent amplitude $\alpha$ \cite{knight87}, and the atom prepared in its upper state $|\varphi(0)\rangle_a = |e\rangle$. The squeezed states not only exhibit reduced noise in one of the quadrature variables compared to the vacuum state, but their photon number distribution, $P_n = |a_n|^2$, also displays a peculiar ringing structure apparent in both the atomic inversion \cite{carmichael89} and the emission spectrum \cite{moyacessa20}. Another interesting case of an initial field state whose nonclassical properties are visibly imprinted in the spectrum is the Schrödinger cat state \cite{dodonov74,vidiella92}, a quantum superposition of two coherent states 
\begin{equation}
	\label{catstate}
	|\Phi(0)\rangle_f = {\cal N} \left(|\alpha\rangle +e^{i\varepsilon} |-\alpha\rangle\right).
\end{equation}
Depending on the value of the relative phase $\varepsilon$, the cat state may exhibit very distinct nonclassical properties, namely, super-Poissonian photon statistics and only even photon numbers for $\varepsilon = 0$ (even coherent state) and sub-Poissonian statistics only odd photon numbers for $\varepsilon = \pi$ (odd coherent state) \cite{vidiella92}. Both states exhibit strong oscillations in their respective photon number distributions. The oscillations in $P_n$ have a direct impact on the spectra for an initial even (odd) coherent state. In Figure \ref{fig:spectrumcat}, we can assess the effect of the oscillations in the photon number distribution on the emission spectrum if we consider the initial state as being $|\Psi(0)\rangle = |e\rangle_a\otimes{\cal N}\left(|\alpha\rangle + |-\alpha\rangle\right)$, (even coherent state). We note the suppression of peaks in the sidebands of the emission spectrum accompanying the missing photon numbers in $P_n$ for this state, in contrast to the case where the field is initially in a coherent state $|\alpha\rangle$, as shown in Figure \ref{fig:spectrumcoherent}. A similar pattern for the spectrum occurs if the field is initially prepared in an odd coherent state, but the peaks are suppressed at different positions in the sidebands. In all figures, the bandwidth of the detector and the atom-field coupling constant will be $\gamma=0.1$ and $\lambda = 5.0$, respectively.


\section{Spectrum for the atomic trapping states}

A remarkable phenomenon occurring in two-level systems is the ``atomic population trapping'', a situation in which the oscillations of the atomic inversion are strongly suppressed. As shown in \cite{zubairy89}, if the atom is initially prepared in a coherent superposition of its energy eigenstates $\left(|g\rangle \ \ \mbox{and} \ \ |e\rangle\right)$ and the field in a coherent state, population trapping occurs as long as certain conditions are met. This behaviour was further investigated in \cite{cirac90}, where the authors found a specific initial state for the atom-field system that leads to perfect trapping, i.e., the atomic inversion remains exactly constant. 

In Section \ref{section2} above, we outlined how quenching of the population inversion, namely the atomic trapping phenomenon, can be understood using the dressed-state coordinates formalism \cite{vidiella99}. Interestingly, even if the initial atom-field state is a product state, the collapse-revival pattern depends not only on the initial coherence of the atom and field by themselves but also on their joint properties as a single quantum system, which in the dressed-state formalism are represented by the function $D_n$ [see Eq.(\ref{dressedness})].

\subsection{Emission spectrum for initial Zaheer-Zubairy trapping states}

The atomic trapping phenomenon in the Jaynes-Cummings model was first discussed in \cite{zubairy89}, where the authors note that the oscillations of the atomic inversion almost disappear if the atom-field system is initially prepared in the state in Equation (\ref{zaheerstate}), that is, the atom in a superposition of its energy eigenstates with a relative phase $-\eta$, and the field in a coherent state $|\alpha\rangle$ with $\alpha = |\alpha|e ^{i\nu_\alpha}$. We have also seen that the condition for atomic trapping considering this state is to have $\eta = \nu_\alpha$ (or $\eta = \nu_\alpha + \pi$). The trapping effect was initially attributed to a ``destructive interference between the atomic dipole and the cavity eigenmode'' \cite{zubairy89}, but the relevance of the dressed states for the population trapping was pointed out in \cite{cirac91}, given that the dressed states have no population inversion. A further step was taken in \cite{vidiella99}, where a quantitative explanation of population trapping based on the dressed-state coordinate formalism was provided. In \cite{zubairy89}, the authors also calculated the emission spectrum and interestingly, they found an asymmetry in the peaks if the initial state is precisely the trapping state in Eq.(\ref{zaheerstate}): the emission in one sideband is suppressed, while the intensity of the remaining sideband is doubled as compared to the case of having the atom initially in the upper state $|e\rangle$. As discussed in \cite{zubairy89}, the emission features of the atom become phase-sensitive, and the spectra will change if the coherent state phase $\nu_\alpha$ is different from the relative phase in the initial atomic superposition state. Additionally, in \cite{cirac91}, the structure of the emission spectrum was briefly addressed from the perspective of the dressed states.

\begin{figure}[ht!]
	\centering\includegraphics[width=9cm]{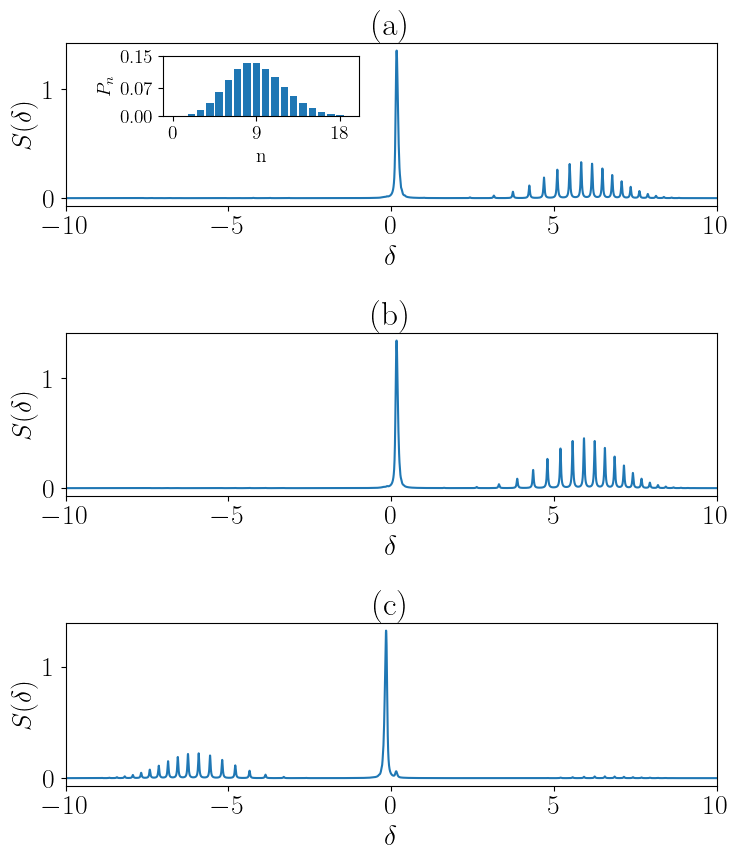}
	\caption{Emission spectrum for the atom initially in a quantum superposition of its energy eigenstates [see Eq.(\ref{zaheerstate})] and the field in a coherent state with $|\alpha| = 3.0$: \hspace{2cm}
		(a) $\Delta = 0$, $\varphi=\frac{\pi}{4}$, $\eta = \nu_\alpha$, \hspace{0.3cm}
		(b) $\Delta = 5.0$, $\tan(2\varphi)=\frac{2\lambda|\alpha|}{\Delta}$, $\eta=\nu_\alpha$, and \hspace{0.3cm}
		(c) $\Delta = 5.0$, $\tan(2\varphi)=\frac{2\lambda|\alpha|}{\Delta}$, $\eta=\pi + \nu_\alpha$. The photon number distribution of the initial state (Poisson distribution) is plotted in the inset of the first graph.}\label{fig:spectrumzsstate01}
\end{figure}

In what follows, we are going to analyse the emission spectrum for initial trapping states using the dressed-state coordinates formalism. We can rewrite the expression for $S(\delta)$ in terms of the coordinates presented above, in a form that is valid for general conditions, including non-product states as
\begin{eqnarray}
	\label{spectrumdressed}
	S(\delta) = \sum_{n=0}^{\infty} w_n^2 \sin^2\left(\frac{\theta_n}{2}\right)\sin^2\beta_n \left[\frac{\gamma \sin^2\beta_{n-1}}{\gamma^2 + \lambda^2(\delta+(\zeta_n+\zeta_{n-1}))^2} + \frac{\gamma \cos^2\beta_{n-1}}{\gamma^2 + \lambda^2(\delta+(\zeta_n-\zeta_{n-1}))^2} \right] \\ \nonumber
	+w_n^2 \cos^2\left(\frac{\theta_n}{2}\right)\cos^2\beta_n \left[\frac{\gamma \sin^2\beta_{n-1}}{\gamma^2 + \lambda^2(\delta-(\zeta_n-\zeta_{n-1}))^2} + \frac{\gamma \cos^2\beta_{n-1}}{\gamma^2 + \lambda^2(\delta-(\zeta_n+\zeta_{n-1}))^2} \right].
\end{eqnarray}
Note that Eq.(\ref{spectrumdressed}) for $S(\delta)$ does not depend on $\chi_n$ and  $\phi_n$, the time-dependent coordinates, given that we are dealing with a time-averaged spectrum. Also, the spectrum in Eq.(\ref{spectrumdressed}) is considerably more compact than the original expression, Eq.(\ref{spectrumbare}). Now it becomes clear the origin of the suppression of the peaks in the emission spectrum. As seen in Eq.(\ref{spectrumdressed}), the intensity of the emitted light as a function of $\delta$ is governed by the weighted dressedness distribution $w_n^2 \sin^2\left(\frac{\theta_n}{2}\right)$. Each one of the sidebands can be associated with a different dressed state, that is, the sideband for $\delta < 0$ is related to $|n-\rangle\,\left(\sin^2\left(\frac{\theta_n}{2}\right)\right)$, and the sideband for $\delta > 0$ is related to $|n+\rangle\,\left(\cos^2\left(\frac{\theta_n}{2}\right)\right)$. In other words, the suppression of the peaks depends on the relative weight of each dressed state in the initial atom-field state, similarly to what happens in the atomic inversion. Conversely, this relation might also allow a direct investigation of a previously unknown initial state through the analysis of the emission spectrum. In Figure \ref{fig:spectrumzsstate01}a we have the spectrum for the Zaheer-Zubairy trapping state with $\eta = \nu_\alpha$, noting that the emission is restricted to $\delta > 0$. On the other hand, if $\eta  = \pi + \nu_\alpha$, the emission occurs for $\delta < 0$. Given that the condition for approximate population trapping found in \cite{vidiella99} (small $D_n$) requires at least one of terms, either $w_n^2 \sin^2\left(\frac{\theta_n}{2} \right)$ or $w_n^2 \cos^2\left(\frac{\theta_n}{2} \right)$ to be small, there can be no two peaks in the spectrum with the same $|\delta|$, and the suppression of one peak implies the doubling of the other, thereby keeping the total emission constant. Here, this is appears as a missing sideband. 

We would also like to analyse the trapping states described in Section \ref{section2} for a non-zero detuning ($\Delta \neq 0$). In this case, the condition to attain the highest possible degree of population trapping is $\tan(2\gamma) \approx 2\lambda\alpha/\Delta$, and slight changes occur in the emission spectrum. We have that the position of each peak in the spectrum is slightly shifted, and an additional asymmetry is created: the intensity of the peaks increases on one side of the spectrum while decreasing on the other, so that emissions closer to the atomic frequency are favoured. If the initial states have no dressing ($\theta_n = \pi/2$), this asymmetry is balanced so that the total emission is constant, as before. However, the total emission might be increased or decreased by combining the effect of the detuning with the populations of the dressed states, as shown in Figure \ref{fig:spectrumzsstate01}. Note that in Figure \ref{fig:spectrumzsstate01}b $(\Delta \neq 0)$, the intensities of the peaks for $\delta > 0$ are higher than those for $\Delta = 0$ in Figure \ref{fig:spectrumzsstate01}a). This is because, for $\Delta > 0$, we have that $\cos^2\beta_n  >  \sin^2\beta_n.$

\subsection{Emission spectrum for initial cat (Yurke-Stoler) trapping states}
\begin{figure}[ht!]
	\centering\includegraphics[width=9cm]{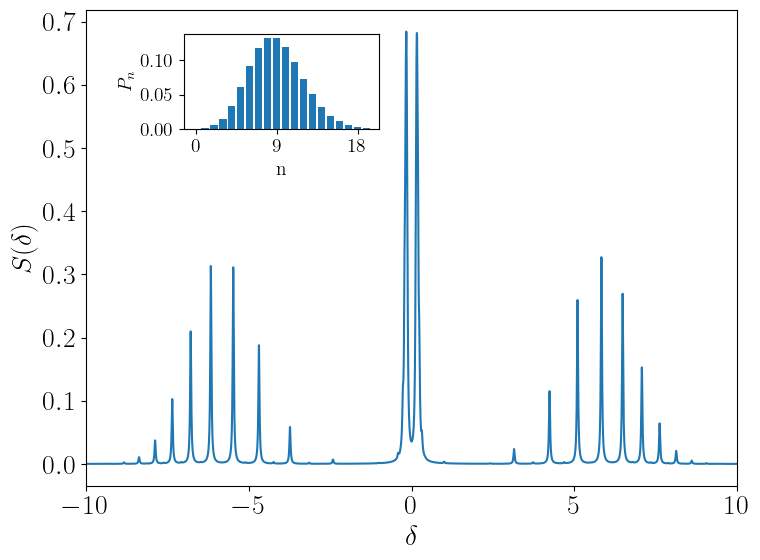}
	\caption{Emission spectrum for the atom initially in a quantum superposition of its energy eigenstates and the field in a Yurke-Stoler state with $|\alpha| = 3.0$ with both signs positive in Eq.(\ref{trappedcatstate}). The photon number distribution for the initial state (Poisson distribution) is plotted in the inset.}\label{fig:spectrumcattrap}
\end{figure}

The method used previously to find the conditions for approximate trapping can also be applied to cases involving more general initial field states. Notably, while trapping is not possible for even or odd cat states, it occurs for initial Yurke-Stoler states of the field, a type of cat state in Eq.(\ref{catstate}) with a relative phase $\epsilon = \pm\pi/2$ \cite{stoler86}. The joint atom-field state is of the form
\begin{equation}
	\label{trappedcatstate}
	|\Psi(0)\rangle = \frac{1}{\sqrt{2}}\left(|e\rangle \pm i e^{-i\nu_\alpha} |g\rangle \right) \otimes {\cal N}\left(|\alpha\rangle \pm i|-\alpha\rangle \right),
\end{equation}
where $\alpha  = |\alpha| e^{i\nu_\alpha}$. We note that the Yurke-Stoler states are nonclassical states having a Poissonian photon number distribution, the same as the coherent states. The fluorescence emission spectrum originating from these states differs significantly from previously obtained spectra, e.g. from initial Zaheer-Zubairy states. Firstly, as seen in Figure \ref{fig:spectrumcattrap} (for $``+"$ signs), the (Yurke-Stoler) spectrum exhibits a triplet, differently from the Zaheer-Zubairy case. A more subtle feature is that there is an asymmetric suppression of peaks; specifically, for $\delta > 0$ ($\delta < 0$) there are only peaks related to even (odd) $n$. Changing any of the $\pm$ signs in Eq.(\ref{trappedcatstate}) results in reflecting $S(\delta)$ across $\delta = 0$. Thus, unlike the case with Zaheer-Zubairy states, these ``cat trapping states'' require an alternating pattern of $|n+\rangle$ or $|n-\rangle$ states to be formed.

\subsection{Emission spectrum for initial perfect trapping states}

\begin{figure}[ht!]
	\centering\includegraphics[width=9cm]{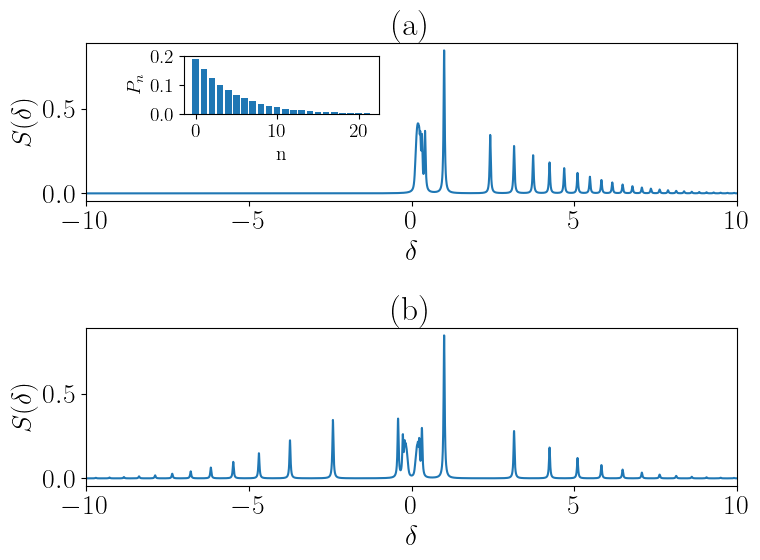}
	\caption{Emission spectrum for perfect trapping states with $z = 0.9$ [see Eqs. (\ref{perfectrapatom}) and (\ref{perfectrapfield}]): (a) $j(n)=1$ and (b) $j(n)=(-1)^{\lfloor n/2 \rfloor}$. The photon number distribution for the initial state (geometrical distribution) is plotted in the inset of the first graph.}\label{fig:spectrumperfecttrap}
\end{figure}

In \cite{cirac90} there were introduced initial product states $|\Psi(0)\rangle = |\varphi(0)\rangle_a\otimes|\Phi(0)\rangle_f$ that lead to a strictly constant atomic inversion, with
\begin{equation}
	\label{perfectrapatom}
	|\varphi(0)\rangle_a = \frac{1}{\left(1 + |z|^2\right)^{1/2}}\left(|g\rangle + z|e\rangle\right),
\end{equation}
and 
\begin{equation}
	\label{phasecoherent}
	|\Phi(0)\rangle_f = \left(1 - |z|^2\right)^{1/2}\sum_{n=0}^\infty z^n |n\rangle.
\end{equation}
The initial field states in Eq.(\ref{phasecoherent}) are the so-called phase-coherent states \cite{shapiro91}. Note that the existence of perfect trapping states is conditioned to having a parameter, $z$  ($|z| < 1$), common to the states of the atom and the field. It was later found \cite{vidiella99} that states of the form
\begin{equation}
	\label{perfectrapfield}
	|\Phi(0)\rangle^{j}_f = \left(1 - |z|^2\right)^{1/2}\sum_{n=0}^\infty j(n) z^n |n\rangle,
\end{equation}
with $j(n) = \pm 1$, also lead to perfect trapping. As expected, the resulting emission spectrum will exhibit an asymmetry similar to the one found in the Zaheer-Zubairy case when the function $j$ is simply $j(n)=1$ (associated with states $|n+\rangle$), with sideband peaks only for $\delta < 0$, as seen in Figure \ref{fig:spectrumperfecttrap}a. On the other hand, the state corresponding to $j(n)=(-1)^n$ (associated with states $|n-\rangle$) results in a (highly asymmetric) spectrum with sideband peaks only for $\delta < 0$. We put forward here a different type of perfect trapping state, for which $j(n) =(-1)^{\lfloor n/2 \rfloor}$. Differently from the above mentioned cases, such state leads to a spectrum having contributions from both $|n+\rangle$ and $|n-\rangle$ states and, therefore, alternated peaks in positive and negative $\delta$, as seen in Figure \ref{fig:spectrumperfecttrap}b. This pattern is similar to the one found in the Yurke-Stoler state case [see Figure \ref{fig:spectrumcattrap}].
%
%
%
%
%
%

\section{Conclusion}

We have explored the emission spectrum in the Jaynes-Cummings model through the lens of the dressed-state coordinate formalism, considering initial states known as ``trapping states''. Under these circumstances the atomic inversion is suppressed but the atomic activity is not, and the two-level atom emits light also in frequencies different from the original field mode. This can bring us a plethora of subtleties which are imprinted in the emitted light. If the atom is initially prepared in its upper energy state, i.e., a non-trapping state, the profile of the emission spectrum can be directly related to the quantum statistical properties of the initial field, reflecting particularities such as oscillations in the photon number distribution \cite{moyacessa20}. However, the situation is different when considering the trapping states. Despite those being product states, specific features of the spectrum like the asymmetry found for the Zaheer-Zubairy trapping states \cite{zubairy89} depend not solely on the photon number distribution, but on the joint properties of the atom-field state. The dressed-state coordinates, as introduced in \cite{vidiella99} constitute an appropriate formalism to address this issue, enabling a direct association between the emission frequencies and the dressed eigenstates of the model. It also allowed us to examine a broader perspective, involving different types of trapping states, such as the ones we introduced here: i) a perfect trapping state that generates a spectrum distinct from those of previously known states, and ii) a new trapping state having the field initially prepared in the Yurke-Stoler cat state, that produces a three-peaked spectrum despite having Poissonian statistics. Those features were successfully explained using the dressed-state coordinates formalism. Furthermore, we addressed a more general situation, the off-resonant case ($\Delta\neq 0$), and obtained simple equations that reveal the effects of detuning and how they, combined with the initial state preparation, can modify the total fluorescence emission of the atom. Our findings illustrate some of the intricate features that can emerge from simple models, such as the Jaynes-Cummmings model, which shows no signs of losing steam even after 60 years of existence.

\section*{Acknowledgments}
A.V.-B. would like to thank the Air Force Office of Scientific Research (AFOSR), USA, under award N${\textsuperscript{\underline{o}}}$ FA9550-24-1-0009, and Conselho Nacional de Desenvolvimento Científico e Tecnol\'ogico, (CNPq), Brazil, via the Instituto Nacional de Ci\^encia e Tecnologia - Informa\c c\~ao Qu\^antica (INCT-IQ), grant N${\textsuperscript{\underline{o}}}$ 465469/2014-0. J.L.T.B. would like to thank the Coordena\c c\~ao de Aperfei\c coamento de Pessoal de N\'ivel Superior, (CAPES), Brazil, via the Programa de Excel\^encia Acad\^emica, (PROEX), grant N${\textsuperscript{\underline{o}}}$ 88887.947492/2024-00.

\bibliographystyle{unsrt}
\bibliography{refsspectrum}



\end{document}